\documentclass{elsart}

\usepackage{epsfig}

\usepackage{amssymb}

\begin{document}

\begin{frontmatter}

\title{Dynamics of condensation of wetting layer in time-dependent Ginzburg-Landau model}

\author{Masao Iwamatsu\thanksref{label2}}
\thanks[label2]{corresponding author E-mail: iwamatsu@ph.ns.musashi-tech.ac.jp, Tel: +81-3-3703-3111 ext.2382, 
Fax: +81-3-5707-2222}

\address{Department of Physics, General Education Center,
Musashi Institute of Technology,
Setagaya-ku, Tokyo 158-8557, Japan }

\begin{abstract}
The dynamics of liquid condensation on a substrate or within a capillary is studied when the wetting film grows via interface-limited growth.  We use a phenomenological time-dependent Ginzburg-Landau (TDGL)-type model with long-range substrate potential.  Using an order parameter, which does not directly represent the density, we can derive an analytic formula for the interfacial growth velocity that is directly related to the substrate potential.  Using this analytic expression the growth of wetting film is shown to conform to a power-law-type growth, which is due to the presence of a long-range dispersion force.  
\end{abstract}

\end{frontmatter}

Liquid condenses on a substrate or within a capillary even if the vapor phase is stable and its pressure is undersaturation. These wetting and capillary condensation are ubiquitous phenomena that have been studied over the centuries.  Although, the equilibrium thermodynamics of those transitions are understood fairly well~\cite{Dietrich,Schick,Deljaguin,Iwamatsu}, the dynamics of condensation are not well studied.  

Recently, partly due to the advances in various types of scanning microscopes~\cite{Butt,Israelachivili}, information concerning the dynamics of condensation of liquid from vapor has become available~\cite{Bocquet,Kohonen,Szoszkiewicz}.  Unfortunately, however, theoretical studies on the dynamics of liquid condensation are quite scarce~\cite{Schmidt,Restagno1,Lum,Leung} and are mostly simulation studies.  In addition, two processes of nucleation and subsequent growth of the wetting layer has not been well separated in previous studies~\cite{Bocquet,Kohonen,Szoszkiewicz}.  

\begin{figure}[htbp]
\begin{center}
\includegraphics[width=0.9\linewidth]{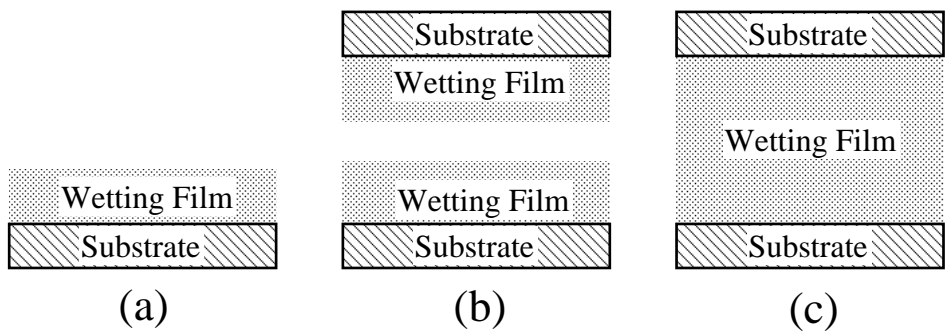}
\end{center}
\caption{
(a) A wetting film on a single substrate, (b)  Two wetting films within a capillary, and (c) the capillary condensation.
}
\label{fig:1}
\end{figure}

In this report, we focus on the growth of the wetting layer after nucleation. We will use the standard time-dependent Ginzburg-Landau (TDGL) model~\cite{Schmidt,Restagno1} to study the dynamics of wetting and capillary condensation from vapor when the interface-limited growth dominates.  In particular, we examine the analytic solution of the growth of wetting film on a single substrate and within a capillary (Fig.~\ref{fig:1}).  The results obtained have a strong connection to the thermodynamic equilibrium properties of wetting and capillary condensation.  Our results predict power law growth of wetting film when the long-range dispersion force represented by the Hamaker constant exists.  The growth velocity is directly expressed by the combination of undersaturated vapor pressure and the Hamaker constant.

We start from the standard TDGL~\cite{Schmidt,Restagno1} model: 
\begin{equation}
\frac{\partial \psi}{\partial t}=-\frac{\delta \mathcal F}{\delta \psi},
\label{eq:1-0}
\end{equation}
where $\delta$ denotes the functional differentiation, $\psi$ is the {\it non-conserved} order parameter as we are interested in condensation phenomena from vapor.  $\mathcal F$ is the free energy functional, which is written as the square-gradient form~\cite{Restagno1}
\begin{equation}
{\mathcal  F}[\psi]=\int \left[\frac{\kappa}{2}(\nabla \psi)^{2}+h(\psi)+V(r)\rho\left(\psi(r)\right)\right]{\rm d}{\bf r}. 
\label{eq:1-1}
\end{equation}
The local part $h(\psi)$ of the free energy functional $\mathcal  F$ determines the bulk vapor-liquid phase diagram and the value of the order parameter $\psi$ in equilibrium phases.  The substrate potential $V(r)$ accounts for the long-range dispersion force from the substrate.  The density $\rho(\psi)$ is given as a function of the order parameter $\psi$, which is assumed to have the following form~\cite{Jou}:
\begin{equation}
\rho(\psi)=6\left(\frac{\psi^2}{2}-\frac{\psi^3}{3}\right).
\label{eq:1-2}
\end{equation}
Therefore the fluid density $\rho$ is indirectly determined from the order parameter $\psi$ in our model.  This somewhat artificial trick allows us to get the analytical expression for the interfacial velocity.

The local part of the free energy (grand potential) $h(\psi)$ we use~\cite{Restagno1} consists of two parts:
\begin{equation}
h(\psi)=h_{0}(\psi)+\Delta \mu \rho\left(\psi\right)
\label{eq:1-3}
\end{equation}
where $\Delta \mu$ is the chemical potential measured from the liquid-vapor coexistence (binodal).  The bulk free energy $h_{0}(\psi)$ is given by
\begin{equation}
h_{0}(\psi) = C\psi^{2}(1-\psi)^{2}. 
\label{eq:1-4}
\end{equation}
where we assume that $\psi_{v}=0$ represents vapor phase, while $\psi_{l}=1$ represents the liquid phase.  Then the densities of liquid and vapor are given by $\rho_{v}=\rho(\psi_{v}=0)=0$ and $\rho_{l}=\rho(\psi_{l}=1)=1$ from eq.~(\ref{eq:1-2}). $C$ is a constant which will be related to the bulk compressibility.  The undersaturated vapor and a metastable liquid phase are characterized by a positive $\Delta \mu>0$, while that of oversaturated vapor is characterized by a negative $\Delta \mu<0$.  Using the ideal gas expression, it is approximately given by
\begin{equation}
\Delta \mu = k_{\rm B}T\ln \frac{p_{\rm sat}}{p_{\rm vap}}
\label{eq:1-4x}
\end{equation}
where $p_{\rm vap}$ is the vapor pressure and $p_{\rm sat}$ is the saturated vapor pressure ($p_{\rm vap}<p_{\rm sat}$). The capillary condensation is expected when $\Delta \mu>0$ and the wetting transition is expected as $\Delta \mu\rightarrow 0^{+}$, while the coexistence of heterogeneous and homogeneous nucleation is predicted when $\Delta \mu<0$~\cite{Iwamatsu2}.  Then the metastable liquid phase and the stable vapor phase in the undersaturated vapor pressure with $\Delta \mu >0$ is characterized by the free energy density (\ref{eq:1-3}) as a function of the order parameter $\psi$.

\begin{figure}[htbp]
\begin{center}
\includegraphics[width=0.7\linewidth]{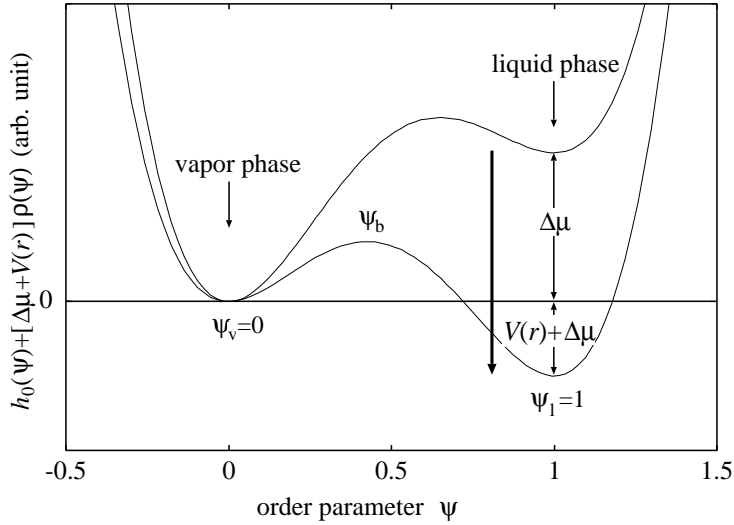}
\end{center}
\caption{The local part of free energy as the function of the order parameter $\psi$ when the effect of the substrate potential $V(r)$ is included.  The substrate potential pushes down (down arrow) the positive free energy of liquid phase to negative stable one.}
\label{fig:2}
\end{figure}

Looking at eqs.(\ref{eq:1-1}) and (\ref{eq:1-3}), the local part of free energy in (\ref{eq:1-1}) can be transformed into the form
\begin{equation}
h(\psi)+V(r)\rho(\psi) = h_{0}(\psi)+\left(\Delta\mu+V(r)\right)\rho(\psi).
\label{eq:1-4y}
\end{equation}
Then, we can interpret the effect of the attractive substrate potential ($V<0$) on the undersaturated vapor is to decrease the chemical potential or the saturation according to
\begin{equation}
\Delta \mu \Rightarrow \Delta\mu+V(r)
\label{eq:1-4z}
\end{equation}
Therefore, the attractive substrate potential turns the positive chemical potential $\Delta\mu$ into the negative local chemical potential $\Delta\mu+V(r)$ (Fig.~\ref{fig:2}). The undersaturated vapor at $\psi_{\rm v}=0$ becomes effectively oversaturated near the substrate, and the attractive substrate potential promotes the nucleation (condensation) of liquid with $\psi_{\rm v}=1$ on the surface of the substrate.  The position of free energy barrier at $\psi_{\rm b}$ depends on the substrate potential $V(r)$ and is given by
\begin{equation}
 \psi_{\rm b}=\frac{1}{2}+\frac{3}{2C}\left(\Delta\mu+V(r)\right).
\label{eq:1-4zz}
\end{equation}  
Furthermore, this negative local chemical potential (\ref{eq:1-4z}) becomes a driving force for the growth of liquid-vapor interface, and thus determines the interfacial velocity.  

The external potential due to the substrate could be modeled by~\cite{Butt,Restagno1}
\begin{equation}
V(x) = A\left(\frac{\sigma}{x+\sigma}\right)^{\alpha}
\label{eq:1-5}
\end{equation}
where $x$ is the distance from the substrate, for the dispersion force~\cite{Israelachivili}. The exponent $\alpha$ is $\alpha=3$ for the three dimensional case.  This liquid-substrate surface tension $A$ is related to the Hamaker constant~\cite{Butt,Israelachivili}.

Now, we consider the thermodynamics of wetting film (Fig.~\ref{fig:1}(a)).  Mean field free energy~\cite{Gartica} for the liquid film of thickness $x$ with constant order parameter $\rho_{l}=\psi_{l}=1$ is given by
\begin{equation}
{\mathcal  F}_{\rm wet}=\left(\Delta \mu x+\int_{0}^{x}V(z)dz+\gamma_{lv}\right)
\label{eq:1-6}
\end{equation}
where $\gamma_{lv}$ is the liquid-vapor surface tension. It can be calculated from the interfacial profile obtained from the Ginzburg-Landau equation $\delta \mathcal F/\delta \psi=0$ for the order parameter $\psi$ at the two-phase coexistence ($\Delta \mu=0$).  

By minimizing this free energy (\ref{eq:1-6}) with respect to the film thickness $x$, we obtain 
\begin{equation}
\Delta\mu+V(x_{min})=0
\label{eq:1-8}
\end{equation}
which gives the thickness $x_{min}$ of the wetting film:
\begin{equation}
x_{min}=\sigma\left(\left(\frac{A}{\Delta \mu}\right)^{1/\alpha}-1\right)
\label{eq:1-9}
\end{equation}
when $A>\Delta \mu$ otherwise $x_{min}=0$ which means no wetting film on the substrate.  On the other hand when $\Delta \mu\rightarrow 0^{+}$ we will have a diverging wetting film thickness $x_{\rm min}\propto \Delta \mu^{-1/3}\rightarrow \infty$~\cite{Dietrich,Schick}.

In capillary (Fig. ~\ref{fig:1} (b)) when the two substrates are separated by the distance $d$, we have to take into account the substrate potential from the opposite substrate. Then the mean-field free energy of two wetting films is given by
\begin{equation}
{\mathcal  F_{\rm film}}(x)=2\left(\Delta \mu x+\int_{0}^{x}\left(V(z)+V(d-z)\right)dz+\gamma_{lv}\right)
\label{eq:1-11}
\end{equation}
Minimization of this free energy give the thickness of wetting film $x_{\rm min}$ from
\begin{equation}
\Delta \mu + V \left( x_{\rm min} \right) + V\left( d-x_{\rm min} \right) =0
\label{eq:1-12}
\end{equation}
similar to (\ref{eq:1-8}) and the free energy ${\mathcal  F_{\rm film}}(x_{\rm min})$ of wetted films, which should be compared with the free energy of capillary-condensed phase (Fig.~\ref{fig:1}(c)):
\begin{equation}
{\mathcal  F_{\rm cc}}=2\left(\Delta \mu \frac{d}{2}+\int_{0}^{d/2}\left(V(z)+V(d-z)\right)dz \right)
\label{eq:1-13}
\end{equation}
with the film thickness $x=d/2$.  The capillary condensation occurs if ${\mathcal F}_{\rm cc}<{\mathcal F}_{\rm film}(x_{\rm min})$~\cite{Gartica}.  

\begin{figure}[htbp]
\begin{center}
\includegraphics[width=0.7\linewidth]{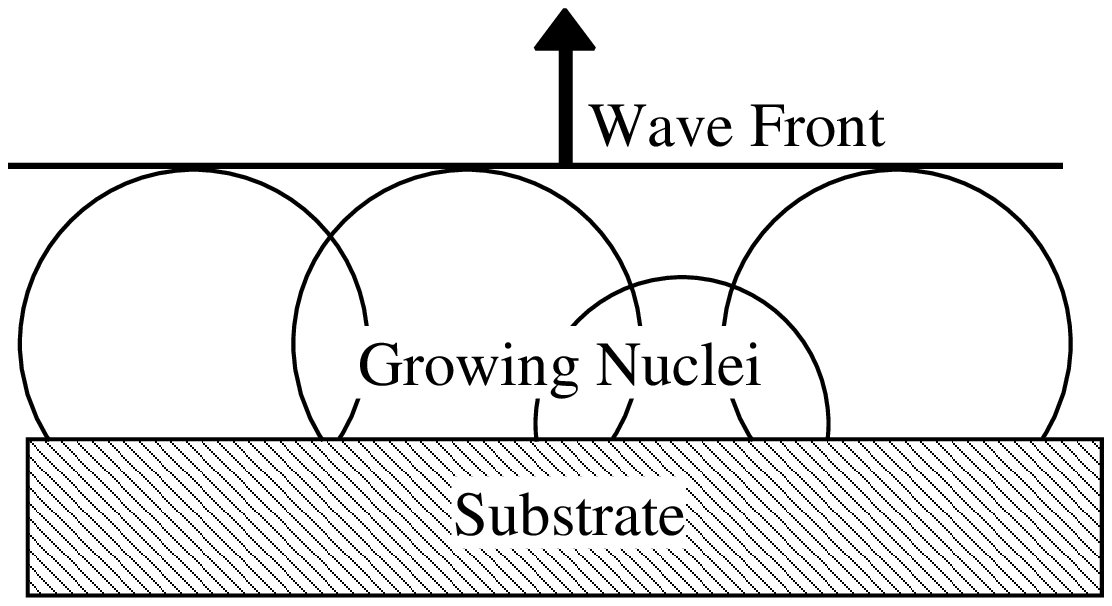}
\end{center}
\caption{
Growing liquid film with a planar wave front as the envelop of spherical wave fronts.
}
\label{fig:3}
\end{figure}

Next, we consider the dynamics of condensation of wetting film using the TDGL model given by (\ref{eq:1-0}).  Since we consider undersaturated vapor phase, the nucleation occurs predominantly on the surface of the substrate.  At a random position on the substrate, the critical nuclei form and grow as a propagating spherical wave.  If the density of nucleus is sufficiently high, eventually, the spherical wave overlap and form a planar wave front from the well-known Huygens's principle (Fig.~\ref{fig:3}). Then, we can consider an effectively one-dimensional problem of planar liquid-vapor interface that is perpendicular to $x$-axis.  The direct variation of the right-hand side of eq.~(\ref{eq:1-0}) using the free energy functional (\ref{eq:1-1})-(\ref{eq:1-4}) gives the Euler-Lagrange equation (\ref{eq:1-0}) in the form
\begin{equation}
\frac{\partial \psi}{\partial t}= \kappa\frac{\partial^{2}\psi}{\partial x^{2}}-\frac{\partial h_{0}}{\partial \psi}-\left(\Delta \mu+V(x)\right)\frac{\partial \rho}{\partial \psi}
\label{eq:1-14a}
\end{equation}
By assuming the solution $\psi(x-R(t))$ around the moving front of wetting film at $R(t)$, equation (\ref{eq:1-14a}) is transformed into~\cite{Chan}
\begin{equation}
\kappa\frac{d^{2}\psi}{dX^{2}}+\dot{R}\frac{d\psi}{dX}-\frac{\partial h_{0}}{\partial \psi}-\left(\Delta\mu+V\left(R(t)+X\right)\right)\frac{\partial \rho}{\partial \psi}=0
\label{eq:1-15}
\end{equation}
with $X=x-R(t)$. Since we are interested in the area around the moving front at $R(t)$, $X$ is of the order of the width of the liquid-vapor interface of moving front where the order parameter $\psi$ changes from $\psi_{v}$ to $\psi_{l}$.  

Now, assuming the traveling wave solution with constant velocity $\dot{R}=v$ and using the approximation
\begin{equation}
V(x)=V(R(t)+X)\simeq V\left(R(t)\right)+\left(\frac{dV}{dx}\right)_{x=R(t)}X \sim V\left(R(t)\right)
\label{eq:1-15a}
\end{equation}
which is valid as long as
\begin{equation}
\frac{1}{V\left(R(t)\right)}\left(\frac{dV}{dx}\right)_{x=R(t)}X<<1
\label{eq:1-15b}
\end{equation}
Then, using eqs.~(\ref{eq:1-2}) and (\ref{eq:1-4}), eq.~(\ref{eq:1-15}) can be transformed into the form
\begin{equation}
\kappa\frac{d^{2}\psi}{dX^{2}}+v\frac{d\psi}{dX}-4C\left(\psi-\psi_{\rm v}\right)\left(\psi-\psi_{\rm b}\right)\left(\psi-\psi_{\rm l}\right)=0
\label{eq:1-15c}
\end{equation}
with $\psi_{\rm v}=0$, $\psi_{\rm l}=1$ and $\psi_{\rm b}$ are the local equilibrium order parameters (Fig.~\ref{fig:2}).  The latter depends on the temporal position $R(t)$ of moving front parametrically through
\begin{equation}
\psi_{\rm b}=\frac{1}{2}+\frac{3}{2C}\left(\Delta\mu+V\left(R(t)\right)\right)
\label{eq:1-15d}
\end{equation}
from (\ref{eq:1-4zz}).

Equation (\ref{eq:1-15c}) can be solved by assuming $d\psi/dX=a\left(\psi-\psi_{\rm v}\right)\left(\psi-\psi_{\rm l}\right)$, which guarantee $d\psi/dX \rightarrow 0$ as $X\rightarrow \pm \infty$ because $\psi\rightarrow \psi_{\rm l}$ as $X\rightarrow -\infty$ and $\psi\rightarrow \psi_{\rm v}$ as $X\rightarrow \infty$.  Inserting this $d\psi/dX$ into eq.~(\ref{eq:1-15c}), and determining the unknown constant $a$, we obtain the velocity of traveling front
\begin{equation}
v=-2\sqrt{\frac{C\kappa}{2}}\left(2\psi_{\rm b}-\left(\psi_{\rm l}+\psi_{\rm v}\right)\right)
\label{eq:15-e}
\end{equation}  
which gives the solution
\begin{equation}
v(R) = -6\sqrt{\frac{\kappa}{2C}}\left(\Delta\mu+V(R)\right).
\label{eq:1-16}
\end{equation}
and the shape of the order parameter around the moving front at $X=0$
\begin{equation}
\psi(X)=\frac{\psi_{l}}{1+\exp\left(\psi_{l}\sqrt{\frac{C}{2\kappa}}X\right)}.
\label{eq:1-16a}
\end{equation}
Equations (\ref{eq:1-16}) and (\ref{eq:1-16a}) can be directly obtained from the results of Chan~\cite{Chan2}.  Now, the front velocity $v$ depend parametrically on the front position $R(t)$, and the width of the interface is $X\sim \sqrt{2\kappa/C}$, which is of the order of molecular size. Then, from eq.~(\ref{eq:1-5}), the condition eq.~(\ref{eq:1-15b})
\begin{equation}
\frac{1}{V\left(R(t)\right)}\left(\frac{dV}{dx}\right)_{x=R(t)}X\sim \frac{X}{R}<<1
\label{eq:1-16b}
\end{equation}
is satisfied as long as the thickness $R(t)$ of wetting film is much larger than the width $X$ of liquid-vapor interface.

Now, the front velocity $v$ depends parametrically on the position $R(t)$ of the front of wetting film.  Consequently, the thickness of the wetting film $x_{min}=R(t\rightarrow \infty)$ is given by the condition where the front velocity vanishes and the growth stops:
\begin{equation}
v(x_{\rm min})=0
\label{eq:1-17}
\end{equation}
which gives eq. (\ref{eq:1-8}) derived from the thermodynamic consideration.

Although the global dynamics of phase transformation should depends on the interface velocity $v(R)$ as well as the nucleation rate $I$ from the general theory of phase transformation~\cite{Iwamatsu2,Christian}, the latter occurs predominantly on the substrate and the nucleation sites will soon be exhausted. Then the wetting dynamics is solely determined from the growth velocity $v(R)$.   A few limiting properties could further be derived from (\ref{eq:1-16}) by replacing the growth velocity $v(R)$ by $dR/dt$. Then we have a differential equation
\begin{equation}
\frac{dR}{dt}= -6\sqrt{\frac{\kappa}{2C}}\left(\Delta\mu+V(R)\right)
\label{eq:1-23}
\end{equation}
with $V(R)$ given by (\ref{eq:1-5}), whose general solution is given by~\cite{Abramobitz}
\begin{equation}
-6\sqrt{\frac{\kappa}{2C}}\Delta\mu\left(t+\mbox{Const.}\right)=(R(t)+\sigma)\left(1- \mbox{$_{2}F_{1}$}\left[\frac{1}{\alpha};1;1+\frac{1}{\alpha};\left(\frac{R(t)+\sigma}{x_{\rm min}+\sigma}\right)\right]\right)
\label{eq:1-24}
\end{equation}
where $_{2}F_{1}$ is the hypergeometric function.

When we approach the binodal $\Delta\mu\rightarrow 0^{+}$, the film thickness $R$ is determined from
\begin{equation}
\frac{dR}{dt}\propto V(R) \propto \frac{1}{R^{\alpha}}
\label{eq:1-18}
\end{equation}
which differs from that derived by Lipowsky~\cite{Lipowsky} for the diffusion-limited growth of wetting film since our model is based on the interface-limited growth.  Equation (\ref{eq:1-18}) gives the power-law growth
\begin{equation}
R(t) \propto t^{1/(\alpha+1)}
\label{eq:1-19}
\end{equation}
for the thickness of wetting film.  A similar power-low growth is predicted from the diffusion-limited growth~\cite{Lipowsky,Burghaus} but the exponent is $1/8$ rather than $1/(\alpha+1)=1/4$.  Here, we used a model of interface-limited growth, and the power-law appears from the power-law long-range force.  In contrast, a similar study of the dynamics of wetting using a TDGL with short-range substrate potential~\cite{Schmidt} gives the logarithmic ($\ln t$) growth of wetting film. The exponent $\alpha$ of the long-range potential (\ref{eq:1-5}) directly gives the growth exponent $1/(\alpha+1)$, and could be used to check the dominant substrate force.

For the wetting film with finite thickness $x_{\rm min}$ with $\Delta \mu>0$, we 
may expand (\ref{eq:1-23}) around the equilibrium thickness $x_{\rm min}$ of the wetting film:
\begin{equation}
\frac{d}{dt}\left(R-x_{\rm min}\right)=6\sqrt{\frac{\kappa}{2C}}\left(\frac{dV}{dx}\right)_{x_{\rm min}}\left(R-x_{\rm min}\right)
\label{eq:1-20}
\end{equation}
where $x_{\rm min}$ is determined from (\ref{eq:1-17}), which gives the exponential relaxation
\begin{equation}
R(t)-x_{min}=\left(x_{0}-x_{\rm min}\right)\exp\left(-t/\tau\right)
\label{eq:1-21}
\end{equation}
toward the equilibrium thickness $x_{\rm min}$, where the relaxation time $\tau$ is given by
\begin{equation}
\tau^{-1}=-6\sqrt{\frac{\kappa}{2C}}\left(\frac{dV}{dx}\right)_{x_{\rm min}}
\propto \frac{\alpha A}{\sigma}\left(\frac{\sigma}{x_{\rm min}}\right)^{\alpha+1}
\label{eq:1-22}
\end{equation}
and $x_{0}$ is the initial value.  The relaxation time $\tau$ is directly proportional to the Hamaker constant $A$.

The growth of wetting film in a closed capillary (Fig.~\ref{fig:1}(b),(c)) can be treated similarly.  The growth velocity is given by
\begin{equation}
v(R) = -6\sqrt{\frac{\kappa}{2C}}\left(\Delta\mu+V(R)+V(d-R)\right)
\label{eq:1-25}
\end{equation}
which leads to the equilibrium wetting film thickness $v(x_{\rm min})=0$ given by eq.~(\ref{eq:1-12}) derived from thermodynamic consideration.  The saturation of the thickness of wetting film is given by eq.~(\ref{eq:1-21}) but with a slightly modified relaxation time $\tau$ 
\begin{equation}
\tau^{-1}=6\sqrt{\frac{\kappa}{2C}}\left[\left(\frac{dV}{dx}\right)_{x_{\rm min}}-\left(\frac{dV}{dx}\right)_{d-x_{\rm min}}\right].
\label{eq:1-26}
\end{equation}
which is again proportional to the Hamaker constant $A$.

In this study, we used a TDGL model~\cite{Restagno1} to study the dynamics of liquid condensation on a substrate or within a capillary when the interface-limited growth dominates. Therefore, our model is more relevant to the liquid condensation from vapor~\cite{Bocquet,Kohonen,Szoszkiewicz} rather the wetting film growth within a binary liquid~\cite{Law,Bonn}. For the latter phase-separation problem, the order parameter must be conserved.  
Therefore diffusion-limited growth is more relevant.  Lipowsky and Huse predicted the power law growth of film thickness $x\propto t^{1/8}$, which was experimentally confirmed recently in cyclohexane-methanol binary mixture~\cite{Bonn}.  

We predicted a similar (fractional) power-law growth $x\propto t^{1/4}$ of wetting film when the interface-limited growth dominates using our TDGL model.  Our model also predicted the exponential saturation of wetting film when the equilibrium thickness is approached. The relaxation time is directly proportional to the Hamaker constant $A$.  

The origin of this power-law time-dependence of wetting film growth is a direct consequence of the same power-law distance-dependence of the dispersion force $V(x)\propto x^{-3}$ from the substrate. Traditionally, the time scale of the growth of wetting film from vapor is usually analyzed only using the classical nucleation theory~\cite{Bocquet,Szoszkiewicz,Restagno1}.  In usual phase transformation, the growth kinetics is determined actually by both the nucleation rate and the growth velocity~\cite{Iwamatsu2,Christian}. In wetting film growth, however, nucleation occurs only on the substrate and the nucleation site will be immediately exhausted.  Then, the phase transformation kinetics of wetting film is characterized by the so-called "site-saturation nucleation"~\cite{Christian,Iwamatsu0}.  Therefore, the nucleation rate plays some role only in the early stage of growth.  Then the time-scale of the growth of wetting film is mainly determined from growth velocity of liquid-vapor interface rather than the nucleation rate of liquid droplets. 

A similar theoretical study of condensation within a capillary using the TDGL model was conducted by Restagno~\cite{Restagno1} et al.  They used TDGL model to analyze the condensation time $\tau$ using the classical nucleation theory.  They showed that the condensation time follows the activation form $\tau \propto \exp(\Delta\Omega/T)$, where $\Delta\Omega$ is the nucleation barrier, and $T$ is the absolute temperature.  Our result for the growth velocity $v$ in Eq.~(\ref{eq:1-16}) does not depend on the temperature and, therefore, contradicts the results of Restagno~\cite{Restagno1} et al. since $\tau\propto 1/v$.  This discrepancy could be due to their arbitrary definition of condensation time $\tau$ even though they stated that their result does not depend sensitively on the precise definition of $\tau$. 

In principle, a direct numerical simulation~\cite{Restagno1,Iwamatsu2} of the long-time behavior of wetting film growth using TDGL model would be possible to confirm the predictions (\ref{eq:1-19}) and (\ref{eq:1-21}).  In practice, however, the confirmation of the power-law behavior (\ref{eq:1-19}) could be difficult even if we use numerically efficient cell dynamics method~\cite{Iwamatsu0} because of the finite size of simulation box.  It seems that there has been no simulation of this kind to study the long-time behavior when the wetting film growth via the interface-limited growth.

So far as the author knows, there has been virtually no detailed experimental study of the dynamics of condensation and wetting film growth from vapor.  Very recently, Kohonen et al.~\cite{Kohonen} for the first time studied the dynamics of condensation within a capillary using surface force apparatus (SFA)~\cite{Israelachivili}.  Their result~\cite{Kohonen} seems qualitatively consistent with the power-law growth $x\propto t^{1/(\alpha+1)}$.  But they analyzed their experimental results only using Langmuir's model~\cite{Langmuir} based on the diffusion limited growth.  They did not find satisfactory agreement. Further detailed analysis of wetting film growth to check the power-law growth, in particular, the exponent $\alpha$ would be interesting.


\begin{thebibliography}{99}
\bibitem{Chan2} There is a typographical error in \cite{Chan}, where eq.(16) should be $v=-\frac{1}{4\gamma}\sqrt{\frac{\kappa}{C}}\left(3\sqrt{9B^2-32AC}+3B\right)$.
\end{thebibliography}
\end{document}